\RequirePackage{amsmath}
\documentclass[12pt]{iopart}
\pdfoutput=1 %to include pdf pictures/figures
\usepackage{graphicx,cite}
\usepackage{amsmath,amsfonts,amssymb}
\usepackage{bm,bbm}
\usepackage[urlcolor=blue]{hyperref}
\hypersetup{colorlinks=true,allcolors=blue}
\usepackage{graphicx}
\usepackage[T1]{fontenc}

\newcommand{\E}[1]{\left\langle #1\right\rangle}
\newcommand{\Es}[1]{\left\langle #1\right\rangle_{\rm s}}

\newcommand{\f}[1]{\mathbf{#1}}
\newcommand{\x}{\f x}
\newcommand{\y}{\f y}
\newcommand{\z}{\f z}

\newcommand{\bsig}{\boldsymbol{\sigma}}

\newcommand{\beps}{{\boldsymbol{\varepsilon}}}
\newcommand{\rj}{\hat{\mathrm j}}
\newcommand{\bj}{\hat{\f j}}
\newcommand{\ps}{p_{\rm s}}%except in operator \bj^s(\x)
\newcommand{\js}{{\f j}_{\rm s}}

\newcommand{\revj}{{\hat{\f j}}^\ddag}
\newcommand{\rrevj}{\hat{\mathrm j}^\ddag}

\renewcommand{\P}[1]{\mathbb{P}\left[#1\right]}

\usepackage{color}\usepackage{xcolor}
\colorlet{mylinkcolor}{blue!66!black!80}
\definecolor{grey}{rgb}{0.6,0.6,.6}
\definecolor{darkgrey}{rgb}{0.4,0.4,.4}
\definecolor{darkgreen}{rgb}{0,0.4,0}
\definecolor{lightgreen}{rgb}{0,0.7,0}
\definecolor{darkred}{rgb}{0.5,0,0}

\begin{document}
\title[{Time-averaged densities and currents with general
    time-dependence}]{On correlations and fluctuations of time-averaged densities and currents with general time-dependence}
\author{Cai Dieball and Alja\v{z} Godec}
\address{Mathematical bioPhysics Group, Max Planck Institute for Multidisciplinary Sciences, 37077 G\"ottingen, Germany}
\ead{agodec@mpinat.mpg.de}

\begin{abstract}
\noindent We present %new
  technical results required for the
description and understanding of 
correlations and fluctuations of the empirical density and current as
well as diverse time-integrated and time-averaged thermodynamic currents
of diffusion processes with a general time dependence 
on all time scales. In particular, we generalize the results from
\textsf{arXiv:2105.10483} (\emph{Phys.\ Rev.\ Lett.\ }, article in press), 
\textsf{arXiv:2204.06553} (\emph{Phys.\ Rev.\ Research}, article in press), and \textsf{arXiv:2206.04034}  
to additive functionals with explicit time dependence and transient
or non-ergodic overdamped diffusion.\ As an illustration we apply
the results to two-dimensional harmonically confined
overdamped diffusion in a rotational flow evolving from a
non-stationary initial distribution.
\end{abstract}
%\maketitle

\section{Introduction}
``Time-average statistical mechanics'' focuses on the study of additive
functionals of stochastic paths and is important in the analysis of
single-particle tracking \cite{Eli_TA,Burov2011PCCP,Ralf}, large deviation
theory
\cite{Touchette2009PR,Chetrite2014AHP,Touchette2018PA,Mallmin2021JPAMT,Coghi2021PRE},
and stochastic thermodynamics
\cite{Seifert2018PA,Koyuk2020PRL,Pietzonka2016PRE,Seifert2005PRL,Pigolotti2017PRL,Seifert2012RPP,Dechant2021PRX,Dechant2021PRR},
to name but a few. The most
important functionals from
a physical point of view
include the ``empirical density''
(also known as local or occupation time)
\cite{Kac1949TAMS,Darling1957TAMS,Aghion2019PRL,Carmi2011PRE,Majumdar2002PRL,Majumdar2002PREa,Majumdar2005CS,Bray2013AP,Bel2005PRL,Lapolla2020PRR}, 
time-integrated and time-averaged currents
\cite{Maes2008PA,Touchette2009PR,Kusuoka2009PTRF,Chetrite2013PRL,Chetrite2014AHP,Barato2015JSP,Hoppenau2016NJP,Touchette2018PA,Mallmin2021JPAMT,Monthus2021JSMTE,Dechant2021PRX,Dechant2021PRR,Hartich2021PRX,Dieball_ARXIV_PRL,Dieball_ARXIV_PRR},
and the time-averaged mean squared displacement (see
e.g.\ \cite{Eli_TA,Burov2011PCCP,Denis,Denis2,Carlos_1,Carlos_2,Carlos_3}).   

Fluctuations of time-averaged observables have a noise floor---they
are bounded from below by the dissipation in a system, which is embodied
within the “thermodynamic uncertainty relation” (TUR)
\cite{Barato2015PRL,Gingrich2016PRL,Gingrich2017JPAMT,Horowitz2019NP,Dechant2018JSMTE,Pietzonka2017PRE,Liu2020PRL,Koyuk2019PRL,Koyuk2018JPAMT,Koyuk2020PRL}.
One may fruitfully exploit this universal 
lower bound on current fluctuations, e.g.\ to gauge the thermodynamic cost of
precision \cite{Barato2015PRL,Cost,Cost_Seifert}, infer dissipation
from fluctuations
\cite{Gingrich2017JPAMT,Horowitz2019NP,Dieball_ARXIV_PRR,Dieball_ARXIV_PRL},
or to derive thermodynamic limits on the temporal extent of anomalous diffusion
\cite{Hartich_TUR}.  

Recent works addressed fluctuations of additive functionals in
transient non-equilibrium systems
\cite{Dechant2018JSMTE,Pietzonka2017PRE,Liu2020PRL}, as well as in
periodically \cite{Koyuk2019PRL,Koyuk2018JPAMT,Koyuk2020PRL} and generally
driven systems  \cite{Koyuk2020PRL}. 

Our aim here is to generalize the direct, stochastic-calculus approach we
developed for steady-state systems in \cite{Dieball_ARXIV_PRR,Dieball_ARXIV_PRL} to transients and systems as well as functionals with explicit time
dependence. Note that this includes non-ergodic systems (see
e.g.\ \cite{Carmi2011PRE,Bel2005PRL}).

The paper is structured as follows. We first set up the formal
background and define the additive functionals in Sec.~\ref{Sec1}. In Section \ref{Sec2} we evaluate the
first moments. In Sec.~\ref{Sec3} we present our main result---a
Lemma that allows a direct evaluation of fluctuations and correlations
of general additive functionals in systems with explicit time
dependence---and derive general results for current fluctuations and
current-density correlations. In Sec.~\ref{Sec4} we illustrate how to
apply the newly developed results by evaluating current-density
correlations  in overdamped diffusion in a rotational flow evolving from a
non-stationary initial distribution. We conclude with a brief
outlook.

%\red{make intro as in JPhysA \cite{Dieball_ARXIV_JPhysA} and PRR,PRL
%  \cite{Dieball_ARXIV_PRR,Dieball_ARXIV_PRL} -- also cite something
%  for transient and non-ergodic and explicit time-dependence
%  \cite{Koyuk2020PRL}. 

%coarse graining PRL and PRR (arxiv) \cite{Dieball_ARXIV_PRR,Dieball_ARXIV_PRL}

%TUR \cite{Barato2015PRL,Gingrich2016PRL}

%TUR for inference \cite{Gingrich2017JPAMT,Horowitz2019NP}

%TUR genralizations:
%transient TUR \cite{Dechant2018JSMTE,Pietzonka2017PRE,Liu2020PRL}, periodically driven \cite{Koyuk2019PRL,Koyuk2018JPAMT}, general driving \cite{Koyuk2020PRL}}

\section{Set-up}\label{Sec1}
Consider overdamped Langevin dynamics with possibly multiplicative
noise and explicit time dependence, described by the
anti-It\^o~(or~H\"anggi-Klimontovich~\cite{anti,anti2}) stochastic
differential equation 
\begin{align}
\rmd\x_\tau=\f F(\x_\tau,\tau)\rmd\tau+\bsig(\x_\tau,\tau)\circledast\rmd\f W_\tau,\label{SDE general} 
\end{align}
with positive definite diffusion matrix $\f
D(\x_\tau,\tau)=\bsig(\x_\tau,\tau)\bsig^T(\x_\tau,\tau)/2$. Assume
that the drift $\f F(\x_\tau,\tau)$ and noise amplitude
$\bsig(\x_\tau,\tau)$ are sufficiently well-behaved for Eq.~\eqref{SDE
  general} to be well-defined with a unique strong solution
(e.g.\ assume that a weak solution exists and $\f F$ and $\bsig$ are
locally Lipschitz continuous \cite{Ikeda1981}).  The anti-It\^o
convention $\circledast\rmd\f W_\tau=\f W_\tau-\f W_{\tau-\rmd\tau}$
is the thermodynamically consistent choice
\cite{Pigolotti2017PRL,Hartich2021PRX,Dieball_ARXIV_PRR}, in
particular it ensures Boltzmann statistics if the drift $\f
F(\x_\tau,\tau)$ is such that the system settles into thermodynamic
equilibrium  \cite{Dieball_ARXIV_PRR}. The time-evolution of the probability density $P(\x,\tau)$ for any initial density $P(\x,\tau=0)$ obeys a Fokker-Planck equation \cite{Risken1989,Gardiner1985}
\begin{align}
\partial_\tau P(\x,\tau)&=[-\nabla_\x\cdot\f F(\x,\tau)+\nabla_\x^T\f
  D(\x,\tau)\nabla_\x]P(\x,\tau)\nonumber\\
&\equiv L(\x,\tau)P(\x,\tau),\label{FPE}
\end{align}
which is equivalent to a continuity equation $[\partial_\tau+\nabla_\x\cdot\bj(\x,\tau)]P(\x,\tau)=0$, with the current operator
\begin{align}
\bj(\x,\tau)\equiv \f F(\x,\tau)-\f D(\x,\tau)\nabla_\x\label{bj} 
\end{align}
that
%(based on the continuity equation)
gives the instantaneous %probability (density)
current as $\f j(\x,\tau)=\bj(\x,\tau)P(\x,\tau)$.

As a special case of Eq.~\eqref{SDE general} we will also study time-homogeneous non-equilibrium steady-state systems, where the stochastic equation of motion reads (curly brackets throughout denote that derivatives only act inside brackets)
\begin{align}
\rmd\x_\tau=\left[\f D(\x_\tau)\{\nabla\log\ps\}(\x_\tau)+\frac{\js(\x_\tau)}{\ps(\x_\tau)}\right]\rmd\tau+\bsig(\x_\tau)\circledast\rmd\f W_\tau,\label{SDE NESS} 
\end{align}
where $\ps$ and $\js$ denote the steady-state density and current
\cite{Dieball_ARXIV_PRR}. Note that (as opposed to
\cite{Dieball_ARXIV_PRR,Dieball_ARXIV_PRL}) we do \emph{not} assume that the
initial distribution is sampled from $\ps$. 

Based on the dynamics defined in Eqs.~\eqref{SDE general} or \eqref{SDE NESS}, we consider time-averaged density and current functionals of the trajectories $[\x_\tau]_{0\le\tau\le t}$ defined as
\begin{align}
\rho_t^V&=\frac{1}{t}\int_0^tV(\x_\tau,\tau)\rmd\tau
\nonumber\\
\f J_t^U&=\frac{1}{t}\int_{\tau=0}^{\tau=t}U(\x_\tau,\tau)\circ\rmd\x_\tau,\label{definitions} 
\end{align}
with $U,V$ differentiable and square integrable functions and $\circ$
denotes the Stratonovich convention of the stochastic integral.  The
density functional $\rho_t^V$ measures the time spent in the region
$V(\x)\ne 0$, weighted by $V(\x)$, while the current $\f J_t^U$
functional measures weighted displacements accumulated in $U$.  In particular, 
for positive $V,U$ that are centered around some point $\x$ and decay on a
finite length scale, one can interpret $\rho_t^V$ and $\f J_t^U$ as the
coarse-grained empirical density and current at $\x$
\cite{Dieball_ARXIV_PRR,Dieball_ARXIV_PRL}.  

In the following, we will derive expressions for the mean values,
correlations and fluctuations of these stochastic quantities and
illustrate them with an example, thereby generalizing the results in
\cite{Dieball_ARXIV_PRR,Dieball_ARXIV_PRL} to non-steady-state initial
conditions and even systems with explicit time-dependence, and thus in
particular also without the existence of a steady state.

\section{First moments}\label{Sec2}
Consider overdamped Langevin dynamics as defined in Eq.~\eqref{SDE
  general} starting from an arbitrary initial density
$P(\x,\tau=0)$. Let $P(\x,\tau)$ be the probability density to find
the particle at position $\x$ after time $\tau$, i.e.\ the solution of
the Fokker-Planck equation in Eq.~\eqref{FPE}. Then the mean value of
the density functional in Eq.~\eqref{definitions} is given by
\begin{align}
\E{\rho_t^V}&=\frac{1}{t}\int_0^t \E{V(\x_\tau,\tau)}\rmd\tau\nonumber\\
&=\frac{1}{t}\int_0^t\rmd\tau\int\rmd^d xV(\x,\tau)P(\x,\tau).\label{mean_rho}
\end{align}
The mean value of the current is in turn given accordingly by
(following closely the approach \cite{Dieball_ARXIV_PRR} using that the It\^o-$\rmd\f W_\tau$-term vanishes on average, integrating by parts, and using $\f D=\f D^T$)
\begin{align}
&\E{\f J_t^U}=\frac{1}{t}\int_0^t\E{U(\x_\tau,\tau)\circ\rmd\x_\tau}
\nonumber\\&
=\frac{1}{t}\int_{\tau=0}^{\tau=t}\E{U(\x_\tau,\tau)\rmd\x_\tau}+\frac{1}{t}\int_{\tau=0}^{\tau=t}\frac{1}{2}\Es{\rmd U(\x_\tau,\tau)\rmd\x_\tau}\nonumber\\
&=\frac{1}{t}\int_0^t\rmd\tau\int\rmd\x P(\x,\tau)\left[U(\x,\tau)\f F(\x,\tau)+\left\{\nabla^T_\x\f D(\x,\tau)\right\}U(\x,\tau)+\f D(\x,\tau)\left\{\nabla_\x U(\x,\tau)\right\}\right]
\nonumber\\&
=\frac{1}{t}\int_0^t\rmd\tau\int\rmd\x P(\x,\tau)\left[U(\x,\tau)\f F(\x,\tau)+\nabla^T_\x\f D(\x,\tau)U(\x,\tau)\right]
\nonumber\\&
=\frac{1}{t}\int_0^t\rmd\tau\int\rmd\x U(\x,\tau)\left[\f F(\x,\tau)-\f D(\x,\tau)\nabla_\x\right]P(\x,\tau)
\nonumber\\&
=\frac{1}{t}\int_0^t\rmd\tau\int\rmd\x U(\x,\tau)\bj(\x,\tau)P(\x,\tau)
\nonumber\\&
=\frac{1}{t}\int_0^t\rmd\tau\int\rmd\x U(\x,\tau)\f j(\x,\tau).\label{mean_j}
\end{align}
The expressions Eq.~\eqref{mean_rho} and \eqref{mean_j} average the
probability density and current over the function $U(\x,\tau)$ and over time $\tau\in[0,t]$,
i.e.\ one can interpret $\rho_t^V$ and $\f J_t^U$ as estimators of
space and time averages of $P(\x,\tau)$ and $\f j(\x,\tau)$. Note that
for time-homogeneous steady-state dynamics  (see Eq.~\eqref{SDE NESS}) these results
are unchanged. They only further simplify for dynamics in Eq.~\eqref{SDE
  NESS} if also the initial condition is sampled from the steady state
$P(\x,\tau=0)=\ps(\x)$, in which case $P(\x,\tau)=\ps(\x)$ and $\f
j(\x,\tau)=\js(\x)$ implies that $\E{\rho_t^V}$ and $\E{\f J_t^U}$ become independent of $t$.

\section{Correlations and fluctuations}\label{Sec3}
We now derive second moments and linear correlations of
the time-averaged density and current in Eq.~\eqref{definitions}. The
derivations for higher moments of currents are more involved than the
first moments but as in \cite{Dieball_ARXIV_PRR} we solve the
complications in the derivation by means of a single Lemma derived in
the following subsection. Note that one could alternatively derive the
following results using a Feynman-Kac approach (and optionally
functional calculus) by appropriately generalizing the approach in
\cite{Dieball_ARXIV_JPhysA}. 

\subsection{Lemma}
In the derivation of expressions for fluctuations and correlations of the
time-averaged quantities we must evaluate correlations
of noise increments $\rmd\f W_\tau$ and functions of
$\x_{\tau'}$. Correlations for $\tau'\le\tau$ vanish by the properties
of the Wiener process. Conversely, correlations for $\tau'>\tau$ are
non-trivial. This problem was solved for steady-state dynamics
in \cite{Dieball_ARXIV_PRR} and via Doob conditioning
\cite{Doob,Chetrite2014AHP,Pigolotti2017PRL} for general
time-homogeneous Langevin systems in the Supplemental Material of
\cite{Dechant2021PRR}. We now generalize the direct approach from
\cite{Dieball_ARXIV_PRR} to overdamped Langevin systems with explicit
time-dependence. 

Consider the $k$-th component $[\bsig(\x_\tau,\tau)\rmd\f W_\tau]_k$ of a
noise increment in an expectation value
$\E{f(\x_\tau,\x_{\tau'},\tau,\tau')\left[\bsig(\x_\tau,\tau)\rmd\f
    W_\tau\right]_k}$ with some (differentiable, square integrable)
function $f$. 
%\begin{align}
%\star=\E{f(\x_\tau,\x_{\tau'},\tau,\tau')\left[\bsig(\x_\tau,\tau)\rmd\f W_\tau\right]_k},\label{expectation term} 
%\end{align}
For $\tau'\le\tau$ this term vanishes due to vanishing correlations
and zero mean of $\rmd\f W_\tau$. Now consider $\tau'>\tau$. 

Given a point $\f x_\tau=\x$ and writing
$\beps\equiv\bsig(\x,\tau)\rmd\f W_\tau$, the equation of motion~\eqref{SDE general} 
rewritten in It\^o form (writing out the anti-It\^o correction term) implies a displacement $\rmd\x_\tau(\x,\tau,\beps)=[\f F(\x,\tau)+\nabla_\x^T\f
  D(\x,\tau)]\rmd\tau+\beps$.  With this we can write the expectation
$\E{f(\cdots)\left[\bsig(\x_\tau,\tau)\rmd\f W_\tau\right]_k}$ as
$\varepsilon_k$ integrated over the probability to be at points
$\x,\x+\rmd\x_\tau(\x,\tau,\beps),\y$ at times
$\tau<\tau+\rmd\tau<\tau'$, i.e. (with joint density
$P(\y,\tau';\x,\tau)$ and conditional density
$P(\y,\tau'|\x,\tau)\equiv P(\y,\tau';\x,\tau)/P(\x,\tau)$; we write
$\mathbbm{1}_{\tau<\tau'}$ for $1$ if $\tau<\tau'$ and $0$ else) 
\begin{align}
&\E{f(\x_\tau,\x_{\tau'},\tau,\tau')\left[\bsig(\x_\tau,\tau)\rmd\f W_\tau\right]_k}\\
&=\mathbbm{1}_{\tau<\tau'}\int\rmd\x\int\rmd\y f(\x,\y,\tau,\tau')\int\rmd\beps\,\mathbb P(\beps)\varepsilon_k P(\y,\tau'|\x+\rmd\x_\tau(\x,\tau,\beps),\tau+\rmd\tau)P(\x,\tau)\,,\nonumber
\end{align}
where the probability $\mathbb P(\beps)$ is given by a Gaussian distribution with zero mean and covariance matrix $2\f D(\x,\tau)\rmd\tau$. Since this distribution
is symmetric around $\f 0$, only terms with even powers of the
components of $\beps$  survive the $\rmd\beps\mathbb P(\beps)$-integration. Note that
\begin{align}
%P(\y,\tau'|\x+\rmd\x_\tau(\x,\tau,\beps),\tau+\rmd\tau)\overset{\rmd\tau\to 0}\longrightarrow
%[1+\rmd\x_\tau(\x,\tau,\beps)\cdot\nabla_\x+\rmd\tau\partial_\tau]P(\y,\tau'|\x,\tau),
P(\y,\tau'|\x+\rmd\x_\tau(\x,\tau,\beps),\tau+\rmd\tau)\overset{\rmd\tau\to 0}\longrightarrow
[1+\rmd\x_\tau(\x,\tau,\beps)\cdot\nabla_\x]P(\y,\tau'|\x,\tau) +\mathcal{O}(\rmd\tau),
\end{align}
and we can neglect the higher orders $\mathcal{O}(\rmd\tau)$ since $\varepsilon_k\mathcal{O}(\rmd\tau)=\mathcal{O}(\rmd\tau^{3/2})$ which (unlike $\varepsilon_k\mathcal{O}(\rmd\tau^{1/2})$) will still give zero after integration in $\tau$. From the zeroth and first order contribution, we see that the only even power of the components of $\beps$ in the above integration gives
\begin{align}
&\E{f(\x_\tau,\x_{\tau'},\tau,\tau')\left[\bsig(\x_\tau,\tau)\rmd\f W_\tau\right]_k}\nonumber\\
&=\mathbbm{1}_{\tau<\tau'}\int\rmd\x\int\rmd\y f(\x,\y,\tau,\tau')P(\x,\tau)\int\rmd\beps\,\mathbb P(\beps)\varepsilon_k\beps\cdot\nabla_\x P(\y,\tau'|\x,\tau),
\end{align}
which, using $\int\rmd\beps\mathbb P(\beps)\varepsilon_k\varepsilon_j=2 D_{kj}(\x,\tau)\rmd\tau$, yields the result for $\tau<\tau'$
\begin{align}
&\E{f(\x_\tau,\x_{\tau'},\tau,\tau')\left[\bsig(\x_\tau,\tau)\rmd\f W_\tau\right]_k}
\nonumber\\&
=\mathbbm{1}_{\tau<\tau'}\rmd\tau\int\rmd\x\int\rmd\y P(\x,\tau)f(\x,\y,\tau,\tau')\left[2\f D(\x,\tau)\nabla_\x P(\y,\tau'|\x,\tau)\right]_k.\label{lemma component}
\end{align}
For scalar products with vector valued functions $\f f$ the result \eqref{lemma component} can be summed over components $f_k$ to obtain
\begin{align}
&\E{\f f(\x_\tau,\x_{\tau'},\tau,\tau')\cdot\bsig(\x_\tau,\tau)\rmd\f W_\tau}
\nonumber\\&
=\mathbbm{1}_{\tau<\tau'}\rmd\tau\int\rmd\x\int\rmd\y P(\x,\tau)\f f(\x,\y,\tau,\tau')\cdot2\f D(\x,\tau)\nabla_\x P(\y,\tau'|\x,\tau).\label{lemma vector}
\end{align}
Eq.~\eqref{lemma vector} is the central result of this work that
allows us to directly deduce expressions for fluctuations and
correlations of densities and currents.
Upon integrating by parts and using symmetry $\f D^T(\x,\tau)=\f D(\x,\tau)$ Eq.~\eqref{lemma vector} could also be rewritten as
\begin{align}
&\E{\f f(\x_\tau,\x_{\tau'},\tau,\tau')\cdot\bsig(\x_\tau,\tau)\rmd\f W_\tau}
\nonumber\\&
=-\mathbbm{1}_{\tau<\tau'}\rmd\tau\int\rmd\x\int\rmd\y P(\y,\tau'|\x,\tau)\nabla_\x\cdot[P(\x,\tau)2\f D(\x,\tau)\f f(\x,\y,\tau,\tau')].
\end{align}

\subsection{Fluctuations and correlations of densities and currents}
Following the developed approach and generalizing the results obtained in
\cite{Dieball_ARXIV_PRR} we now derive expressions for fluctuations
and correlations of densities and currents for arbitrary initial
conditions. 

For two time-averaged densities $\rho_t^U,\rho_t^V$, the covariance (variance for $U=V$) is given by
\begin{align}
&\E{\rho_t^U\rho_t^V}-\E{\rho_t^U}\E{\rho_t^V}=t^{-2}\int_0^t\rmd\tau\int_0^t\rmd\tau'\left[\E{U(\x_\tau,\tau)V(\x_{\tau'},\tau')-\E{U(\x_\tau,\tau)}\E{V(\x_{\tau'},\tau')}}\right]
\nonumber\\
&\qquad=t^{-2}\int_0^t\rmd\tau\int_0^t\rmd\tau'\int\rmd\x\int\rmd\y U(\x,\tau)V(\y,\tau')\left[P(\x,\tau;\y,\tau')-P(\x,\tau)P(\y,\tau')\right].\label{density covar}
\end{align} 
Note that this result can be interpreted as correlations caused by
differences of $P(\x,\tau;\y,\tau')$ and $P(\x,\tau)P(\y,\tau')$,
averaged over time and over functions $U,V$.  More precisely, the
two-point function $P(\x,\tau;\y,\tau')$ can be understood to be
characterized by all paths with $\x_\tau=\x$ and $\x_{\tau'}=\y$. For
further interpretation, in particular for the case of steady-state
dynamics, see \cite{Dieball_ARXIV_PRR,Dieball_ARXIV_PRL}. 

For the correlation of $\f J_t^U$ and $\rho_t^V$ we first consider the
expectation of the product and carry out the same steps as in
Eq.~\eqref{mean_j}, 
\begin{align}
&t^2\E{\f J_t^U\rho_t^V}
=\int_0^t\rmd\tau'\int_{\tau=0}^{\tau=t}\E{U(\x_\tau,\tau)\circ\rmd\x_\tau V(\x_{\tau'},\tau')}\nonumber\\
&=\int_0^t\rmd\tau\int_0^t\rmd\tau'\int\rmd\x\int\rmd\y U(\x,\tau)V(\y,\tau')\bj(\x,\tau)P(\y,\tau';\x,\tau)\nonumber\\
&\quad
+\int_0^t\rmd\tau'\int_{\tau=0}^{\tau=t}\E{U(\x_\tau,\tau)\bsig(\x_\tau,\tau)\rmd\f W_\tau V(\x_{\tau'},\tau')}.\label{correlation ansatz}
\end{align}
Comparing with the calculation in Eq.~\eqref{mean_j}, the noise term
no longer vanishes since terms with $\tau<\tau'$ give non-trivial
correlations according to Eq.~\eqref{lemma component}, which in turn gives 
\begin{align}
&\int_0^t\rmd\tau'\int_{\tau=0}^{\tau=t}\E{U(\x_\tau,\tau)\bsig(\x_\tau,\tau)\rmd\f W_\tau V(\x_{\tau'},\tau')}
=\int_0^t\rmd\tau'\int_0^t\rmd\tau\mathbbm{1}_{\tau<\tau'}\times\nonumber\\
&\int\rmd\x\int\rmd\y U(\x,\tau)V(\y,\tau')\left[2P(\x,\tau)\f D(\x,\tau)\nabla_\x P(\x,\tau)^{-1}\right]P(\y,\tau';\x,\tau),
\end{align}
where we rewrote $P(\y,\tau'|\x,\tau)=P(\x,\tau)^{-1}P(\y,\tau';\x,\tau)$. Introducing the adapted current operator 
\begin{align}
\revj(\x,\tau)\equiv \bj(\x,\tau)+2P(\x,\tau)\f D(\x,\tau)\nabla_\x P(\x,\tau)^{-1},\label{revj} 
\end{align}
%and rewriting the $\tau,\tau'$-integrations with $t_1\equiv\min(\tau,\tau')\le t_2\equiv\max(\tau,\tau')$
we thus obtain from Eq.~\eqref{correlation ansatz} an expression for
the current-density correlation that reads 
\begin{align}
&\E{\f J_t^U\rho_t^V}-\E{\f J_t^U}\E{\rho_t^V}
=t^{-2}\int_0^t\rmd\tau\int_0^t\rmd\tau'\int\rmd\x\int\rmd\y U(\x,\tau)V(\y,\tau')\times\nonumber\\
&\qquad\left[\mathbbm{1}_{\tau>\tau'}\bj(\x,\tau)+\mathbbm{1}_{\tau<\tau'}\revj(\x,\tau)\right]\left[P(\y,\tau';\x,\tau)-P(\x,\tau)P(\y,\tau')\right].\label{current density correlation} 
\end{align} 
Note that to write the expression more compactly, we used that
$\bj(\x,\tau)P(\x,\tau)=\revj(\x,\tau)P(\x,\tau)=\f j(\x,\tau)$.  For symmetry reasons and since the difference vanishes, we wrote $\mathbbm{1}_{\tau>\tau'}$ instead of $\mathbbm{1}_{\tau\ge\tau'}$. 

The expression \eqref{current density correlation} is a natural generalization of Eq.~\eqref{density
  covar} with the current operators $\bj,\revj$ appearing. Recall that
$\bj$ is the current operator entering the Fokker-Planck equation, see
Eqs.~\eqref{FPE}-\eqref{bj}. The adapted operator $\revj$ defined in
Eq.~\eqref{revj} accounts for the fact that trajectories contributing
to $P(\y,\tau';\x,\tau)$ that first visit $\x$ and later $\y$
(i.e.\ $\tau<\tau'$) have, compared to the Fokker-Planck evolution,
altered statistics, since displacements at $\x$ correlate with
probabilities of reaching $\y$ later.  For the particular case of
steady-state systems, the special case of the correlation result~\eqref{current density correlation} and the adapted current
operator were discussed in detail, and explained using a generalized
time-reversal symmetry, in 
\cite{Dieball_ARXIV_PRR,Dieball_ARXIV_PRL}.  Note that for the case of
time-homogeneous dynamics (in particular steady-state dynamics defined in
Eq.~\eqref{SDE NESS}), the Fokker-Planck current operator
Eq.~\eqref{bj} does not have an explicit time dependence such that
$\bj(\x,\tau)$ in Eq.~\eqref{current density correlation} simplifies
to $\bj(\x)$. However, the adapted current operator $\revj(\x,\tau)$
defined in Eq.~\eqref{revj} retains explicit time-dependence even for
time-homogeneous dynamics.  Only in the case of steady-state systems
with steady-state initial conditions (where $P(\x,\tau)=\ps(\x)$ for
all $\tau$) $\revj$ has no explicit time dependence, and simplifies to
the negative $\bj$ with inverted steady-state current $\js\to-\js$
\cite{Dieball_ARXIV_PRR,Dieball_ARXIV_PRL}.  

Covariances of components $m,n$ of time-integrated currents $J_{t,m}^U$ and $J_{t,n}^V$ can be obtained analogously by considering
\begin{align}
t^2\E{J_{t,m}^U J_{t,n}^V}
=\int_{\tau'=0}^{\tau'=t}\int_{\tau=0}^{\tau=t}\E{U(\x_\tau,\tau)\circ\rmd x^m_\tau V(\x_{\tau'},\tau')\circ\rmd x^n_{\tau'}},\label{current ansatz}
\end{align}
where both $\circ\rmd\x_t$ increments split into $\rmd t$ and $\rmd\f
W_t$ terms. The $\rmd\tau\rmd\tau'$ terms give rise to the current
operator $\bj$ as in Eqs.~\eqref{mean_j},\eqref{correlation ansatz},
but now its components $\rj_m(\x,\tau)$ and $\rj_n(\y,\tau')$
appear. The $\rmd W_\tau\rmd W_{\tau'}$ term yields (by It\^o's
isometry, i.e.\ ``delta-correlated white noise'')
\begin{align}
&\int_{\tau'=0}^{\tau'=t}\int_{\tau=0}^{\tau=t}\E{U(\x_\tau,\tau)\left[\bsig(\x_\tau,\tau)\rmd\f W_\tau\right]_m V(\x_{\tau'},\tau')\left[\bsig(\x_{\tau'},\tau')\rmd\f W_{\tau'}\right]_n}\nonumber\\
&=\int_0^t\E{U(\x_\tau,\tau)V(\x_\tau,\tau)2D_{mn}(\x_\tau)}\rmd\tau\nonumber\\
&=2\int_0^t\rmd\tau\int\rmd\x U(\x,\tau)V(\x,\tau)D_{mn}(\x)P(\x,\tau).
\end{align}
The mixed term $\rmd\tau'\rmd\f W_\tau$ (and equivalently
$\rmd\tau\rmd\f W_{\tau'}$) in Eq.~\eqref{current ansatz} according to
calculations as in Eq.~\eqref{mean_j} and using Eq.~\eqref{lemma
  component} gives 
\begin{align}
&\int_0^t\rmd\tau'\!\!\int_{\tau=0}^{\tau=t}\!\!\!\!\E{U(\x_\tau,\tau)\left[\bsig(\x_\tau,\tau)\rmd\f W_\tau\right]_m\left[V(\x_{\tau'},\tau')\f F(\x_{\tau'},\tau')+\left\{\nabla \f D V\right\}\!(\x_{\tau'},\tau')\right]_n}
=\mathbbm{1}_{\tau<\tau'}\times
\nonumber\\&
\int_0^t\rmd\tau\int_0^t\rmd\tau'\int\rmd\x\int\rmd\y U(\x,\tau)V(\y,\tau')\rj_n(\y,\tau')\left[2P(\x,\tau)\nabla_\x P(\x,\tau)^{-1}\right]_m P(\y,\tau';\x,\tau).
\end{align}
Collecting all terms and using the and notation $\rrevj_{m}$ for the components of $\revj$ in Eq.~\eqref{revj}, 
%\red{and using the definition \eqref{revj}, i.e.\  \begin{align}
%\cdots&=\rj_m(\x,\tau)\rj_n(\y,\tau')-\mathbbm{1}_{\tau<\tau'}\rj_n(\y,\tau')\left[2P(\x,\tau)\nabla_\x P(\x,\tau)^{-1}_m\right]-\mathbbm{1}_{\tau>\tau'}\rj_m(\x,\tau)\left[2P(\y,\tau')\nabla_\y P(\y,\tau')^{-1}_n\right]\nonumber\\
%&=\mathbbm{1}_{\tau<\tau'}\rrevj_m(\x,\tau)\rj_n(\y,\tau')
%+\mathbbm{1}_{\tau>\tau'}\rj_m(\x,\tau)\rrevj_n(\y,\tau')
%\end{align}}
we obtain for Eq.~\eqref{current ansatz}
\begin{align}
&t^2\E{J_{t,m}^U J_{t,n}^V}=2\int_0^t\!\!\rmd\tau\!\int\!\!\rmd\x
  U(\x,\tau)V(\x,\tau)D_{mn}(\x,\tau)P(\x,\tau)+\int_0^t\!\!\rmd\tau\!\!\int_0^t\!\!\rmd\tau'\!\!
  \int\!\!\rmd\x\! \int\!\!\rmd\y
\times\nonumber\\&
U(\x,\tau)V(\y,\tau')\left[\mathbbm{1}_{\tau<\tau'}\rrevj_m(\x,\tau)\rj_n(\y,\tau')+\mathbbm{1}_{\tau>\tau'}\rj_m(\x,\tau)\rrevj_n(\y,\tau')\right]P(\y,\tau';\x,\tau).\label{current covariance} 
\end{align}
From the derivation one sees that the first term is the
$\tau=\tau'$-contribution (see also \cite{Dieball_ARXIV_PRR}). This is
the natural generalization of the results in Eqs.~\eqref{density covar}
and \eqref{current density correlation}, with the interpretation of
non-trivial displacements (and thus $\revj$ instead of $\bj$) for
currents evaluated at earlier times (see above and
\cite{Dieball_ARXIV_PRR,Dieball_ARXIV_PRL}).  As before, for
time-homogeneous dynamics $\bj(\x,\tau)$ simplifies to $\bj(\x)$ and
in the special case of steady-state dynamics (see Eq.~\eqref{SDE NESS})
with steady-state initial conditions, $\revj(\x,\tau)$ simplifies to
$\revj(\x)$. This special case was discussed
and explained using generalized time-reversal symmetry in
\cite{Dieball_ARXIV_PRR,Dieball_ARXIV_PRL}.% for the particular case of non-equilibrium steady-states. 

\section{Example}\label{Sec4}
To present a concrete minimal example, we consider a two-dimensional
harmonically confined overdamped diffusion in a rotational flow
(i.e.\ an irreversible Ornstein-Uhlenbeck process)
\begin{align}
\rmd\x_t=-\begin{bmatrix}1&-\Omega\\\Omega&1\end{bmatrix}\x\rmd t+\sqrt{2}\rmd\f W_t.\label{OUP}
\end{align}
Assuming that the initial density $P(\x,\tau=0)$ is Gaussian, the
solution $P(\x,\tau)$ of the Fokker-Planck equation corresponding to
Eq.~\eqref{OUP} is well known to be a Gaussian density for all
$\tau\ge0$ (see e.g.\ \cite{Gardiner1985}). We choose $U$ to be a two-dimensional Gaussian centered at $\z$ with width $h$, i.e.
\begin{align}
U_\z(\x)=\frac{1}{2\pi h^2}\exp\left[-\frac{(\x-\z)^2}{2h^2}\right].\label{Gauss window}
\end{align}
Due to the Gaussianity of $P(\x,\tau)$ and $U_\z(\x)$, all spatial
integrals entering the results Eqs.~\eqref{density covar},
\eqref{current density correlation} and \eqref{current covariance} can
be performed analytically, e.g.\ using the computer algebra system
\textsf{SymPy} \cite{SymPy} (as outlined in the Supplemental Material
of \cite{Dieball_ARXIV_PRL}).  The two remaining time-integrals are
computed numerically.  For simplicity we only consider the (non-steady-state) initial
condition in a point, i.e.\ $P(\x,\tau=0)=\delta(\x-\x_0)$. For this
initial condition, via a left-right decomposition for the process
Eq.~\eqref{OUP} (see e.g.\ \cite{Risken1989}) or by solving the Lyapunov equation, we have the time-dependent density
\begin{align}
P(\x,\tau)&=\frac{1}{2\pi(1-\rme^{-2\tau})}\exp\left[\frac{-\left (\x-\rme^{-\tau}\begin{bmatrix}\cos(\Omega \tau)&\sin(\Omega \tau)\\-\sin(\Omega \tau)&\cos(\Omega \tau)\end{bmatrix}\x_0\right )^2}{2(1-\rme^{-2\tau})\vphantom{\begin{bmatrix}0\\0\end{bmatrix}}}\right],
\label{density OUP}
\end{align}
i.e.\ the mean value
$\E{\x_\tau}=\rme^{-\tau}\begin{bmatrix}\cos(\Omega \tau)&\sin(\Omega
  \tau)\\-\sin(\Omega \tau)&\cos(\Omega \tau)\end{bmatrix}\x_0$ moves
on a spiral shape towards the center.  The case $\Omega=0$ corresponds
to the equilibrium process, i.e.\ harmonically confined overdamped diffusion without rotational flow.

\begin{figure}
\centering
\includegraphics[width=.95\textwidth]{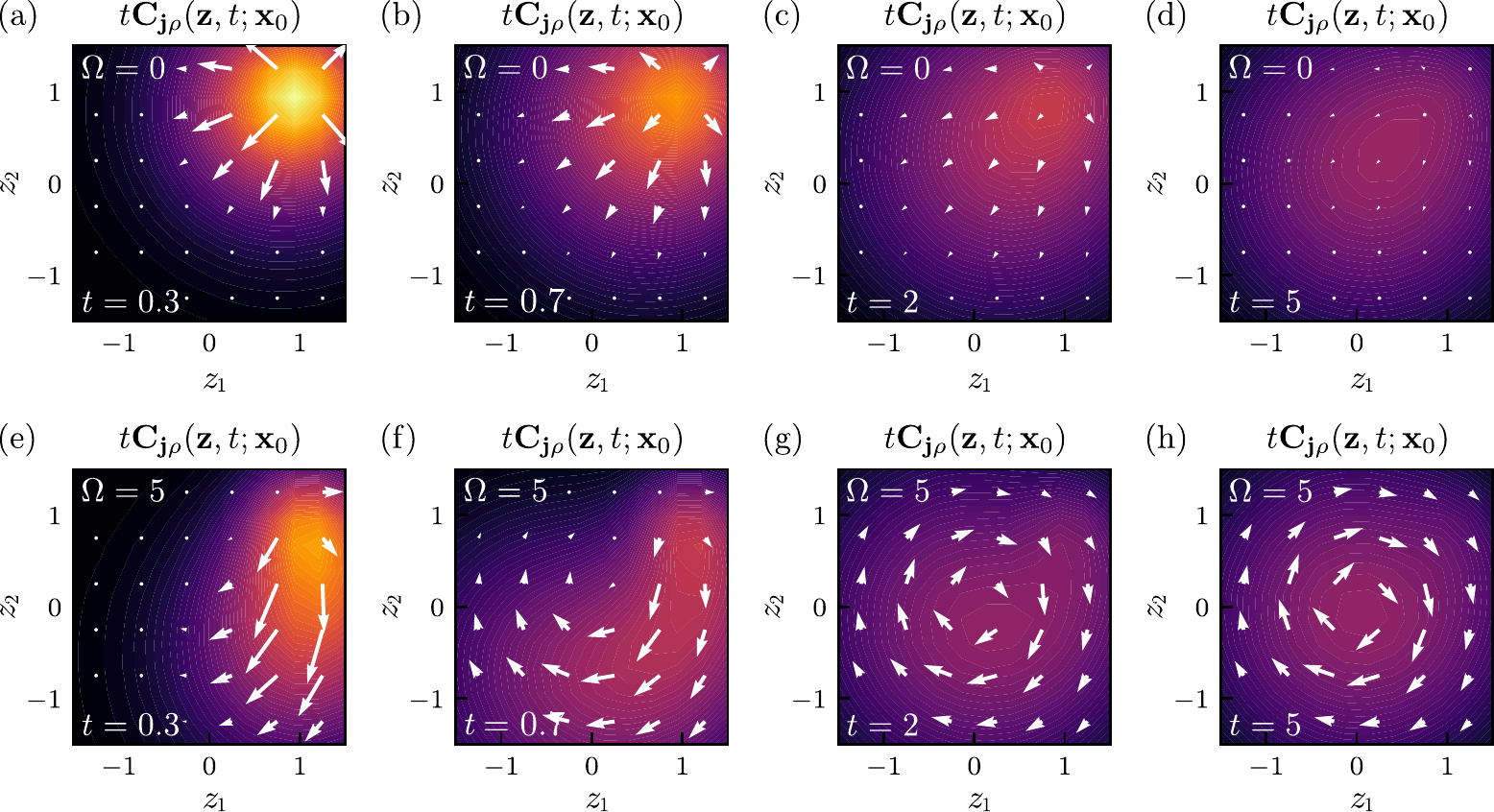}
\caption{White arrows depict the correlation result multiplied by
  time, $t\f C_{\f j\rho}(\z,t;\x_0)$ as in Eq.~\eqref{current density
    correlation} with $\x_0=(1,1)^T$ for the process in Eq.~\eqref{OUP} with $\Omega=0$ in
  (a-d), $\Omega=5$ in (e-h), and $U_\z$ as in Eq.~\eqref{Gauss
    window}. The position $\z=(z_1,z_2)^T$ around which the
  correlation is evaluated varies along the respective axes.  The
  color gradient depicts the mean time-averaged density $\E{\rho_t^{U_\z}}$, i.e.\ the time spent around $\z$ weighted by $U_\z$. Time increases from left (a,e) to right (d,h), $t=0.3,\,0.7,\,2,\,5$
% 300 points for each numerical time-integration
\label{fg1}}
\end{figure}

For this example, we compute the density-current correlation vector as in Eq.~\eqref{current density correlation},
\begin{align}
&\f C_{\f j\rho}(\z,t;\x_0)\equiv\E{\f J_t^{U_\z}\rho_t^{U_\z}}_{\x_0}-\E{\f J_t^{U_\z}}_{\x_0}\E{\rho_t^{U_\z}}_{\x_0}=t^{-2}\int_0^t\rmd\tau\int_0^t\rmd\tau'\int\rmd\x\int\rmd\y\times\nonumber\\
&\quad U_\z(\x)U_\z(\y)\left[\mathbbm{1}_{\tau>\tau'}\bj(\x)+\mathbbm{1}_{\tau<\tau'}\revj(\x,\tau)\right]\left[P(\y,\tau';\x,\tau)-P(\x,\tau)P(\y,\tau')\right].\label{current density correlation re-stated} 
\end{align}
with Gaussian $U_\z$ as in Eq.~\eqref{Gauss window}.  In
Fig.~\ref{fg1} we show the time evolution and spatial dependence of
this correlation vector. For long times without driving $\Omega=0$, we
see that $t\f C_{\f j\rho}(\z,t;\x_0)\to 0$. This corresponds to the
limit when the initial condition is forgotten, i.e.\ for long times
$\f C_{\f j\rho}(\z,t;\x_0)$ approaches the result of $\f C_{\f
  j\rho}(\z,t)$ for steady-state initial conditions where in
equilibrium ($\Omega=0$) we have
$P(\y,\tau';\x,\tau)=P(\x,\tau';\y,\tau)$ (time-reversal symmetry) and
$\revj(\x)=-\bj(\x)$ \cite{Dieball_ARXIV_PRR}, implying $\f C_{\f
  j\rho}(\z,t)=\f 0$ at all $\z$.  In the case $\Omega\ne 0$, the
correlation $t\f C_{\f j\rho}(\z,t;\x_0)$ becomes constant for long
times, where $\f C_{\f j\rho}\propto t^{-1}$ represents the large-deviation limit of the correlation result, which agrees with the large-deviation limit for the process starting in steady-state initial
conditions \cite{Dieball_ARXIV_PRR}. This has a spatial dependence
similar to the steady-state current $\js(\z)$ but averaged over the
Gaussian $U_\z$.  By comparison with the color gradient we see in all
panels in Fig.~\ref{fg1}, as expected, that large values of the
correlation can only occur at points that are visited for a
significant amount of time, i.e.\ with not too small $\E{\rho_t^{U_\z}}$. 

\begin{figure}
\centering
\includegraphics[width=1.\textwidth]{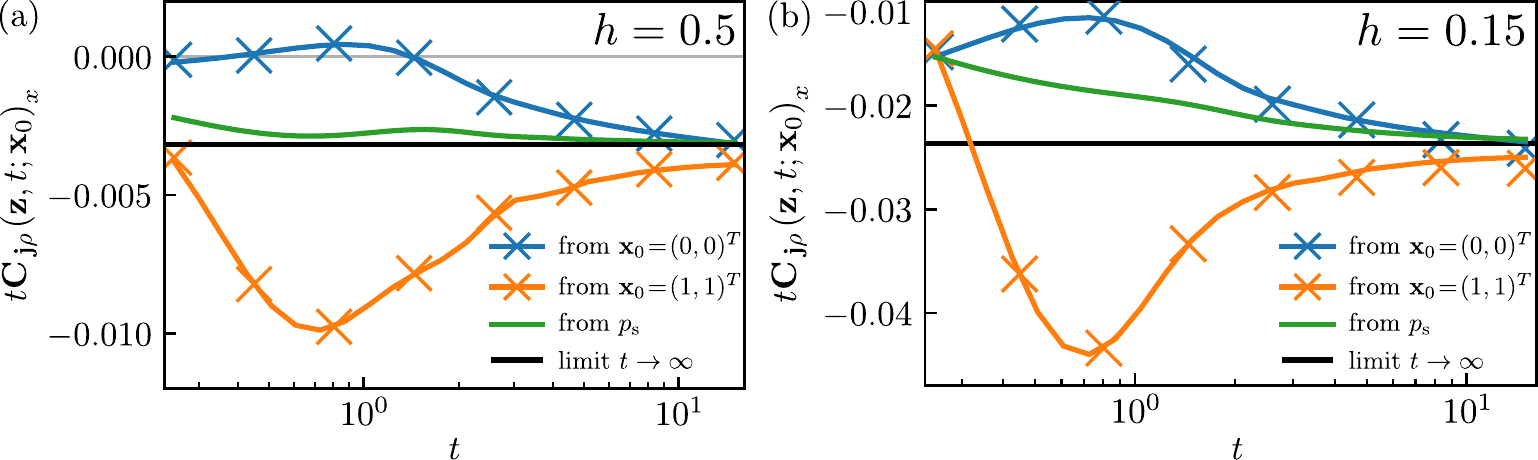}
\caption{Quantitative depiction of the time-dependence of the $x$-component of the current-density correlation $t{\f C_{\mathbf j\rho}(\mathbf{z},t;\mathbf{x}_0)}_x$ with $\mathbf{z}=(0,-0.2)^T$ for the process in Eq.~\eqref{OUP} with $\Omega=3$ and $U_\z$ as in Eq.~\eqref{Gauss
    window} with (a) $h=0.5$ and (b) $h=0.15$ for different initial conditions (colors). The new analytical result (blue and orange lines; Eq.~\eqref{current density correlation}) is confirmed by simulations (crosses; for each $t$, (a) $10^5$ and (b) $10^6$ trajectories with $10^3$ time-steps each were simulated according to the stochastic Euler algorithm). For $t\to\infty$, irrespective of the initial condition, all result approach the same large-deviation limit.}
\label{fg2}
\end{figure}

In addition to the qualitative behavior shown in Fig.~\ref{fg1}, we
present a quantitative evaluation of the correlation result multiplied
by time, $t\f C_{\f j\rho}(\z,t;\x_0)$, for a single $\z$ in
Fig.~\ref{fg2}.  Simulations shown in Fig.~\ref{fg2} confirm the
theoretical result in Eq.~\eqref{current density correlation}
(re-stated in Eq.~\eqref{current density correlation re-stated}).  As
mentioned above this result approaches the large-deviation limit for
long times.  Moreover, for long times the initial condition will
become irrelevant, i.e.\ $t\f C_{\f j\rho}(\z,t;\x_0)$ approaches the
result for $t\f C_{\f j\rho}(\z,t)$ for steady-state initial
conditions \cite{Dieball_ARXIV_PRR}.  First note that, due to the
time-integration, deviations for short times are only slowly
'forgotten' with order $t^{-1}$ (instead of exponentially fast with
some Poincar\'e time scale).  Interestingly, we see in Fig.~\ref{fg2}a
that for substantial coarse-graining (i.e.\ rather large $h=0.5$ in
$U_\z$ in Eq.~\eqref{Gauss window}), the result for $t\f C_{\f
  j\rho}(\z,t;\x_0)$ starting in a point only approaches the
corresponding value for steady-state initial condition (green curve)
in the large deviation regime (black line), but \emph{not} before.
Going to smaller coarse graining $h=0.15$ in Fig.~\ref{fg2}b, we see
that the process starting in the center $\x_0=(0,0)^T$ (blue line)
features the arguably more intuitive behavior, by first approaching
the steady-state (green line) and later the large deviation result
(black line).  However, for a different initial condition (see orange
line)  the steady state curve is again only approached in the large
deviation limit. This highlights the long-lasting and non-trivial
effects of the time-integration and underscores why interpreting
time-average observables, in particular those involving currents,
remains challenging.

\section{Conclusion}
To summarize, we presented a new Lemma \eqref{lemma vector} that
enabled us to
derive results Eqs.~\eqref{density covar},
\eqref{current density correlation} and \eqref{current covariance} for
correlations and fluctuations of the time-averaged density and current
Eq.~\eqref{definitions} for general Langevin dynamics defined in Eq.~\eqref{SDE
  general} with general initial conditions.  This generalization of
the recent results derived for non-equilibrium steady states
\cite{Dieball_ARXIV_PRR,Dieball_ARXIV_PRL} may improve the
understanding of inference of densities and currents with the
estimators $\rho_t^V$ and $\f J_t^U$ (in particular in connection with
the notion of coarse graining
\cite{Dieball_ARXIV_PRR,Dieball_ARXIV_PRL}) in cases where the
dynamics does \emph{not} evolve from the steady-state, or is not
time-homogeneous.  Importantly, the strategy of inferring dissipation from the
current variance (see Eq.~\eqref{current covariance}) via the thermodynamic
uncertainty relation (TUR)
\cite{Barato2015PRL,Gingrich2016PRL,Horowitz2019NP,Gingrich2017JPAMT}
remains valid. Generalized versions of the TUR, e.g.\ for general initial
conditions \cite{Dechant2018JSMTE} or time-dependent dynamics
\cite{Koyuk2020PRL}, already exist.  A recently improved version of the
TUR that includes current-density correlations (see Eq.~\eqref{current density
  correlation}) is,  however,  so far only available for steady-state
systems with steady-state initial conditions
\cite{Dechant2021PRX}. Notably, as we will show in a forthcoming
publication, Lemma \eqref{lemma
  vector} allows the correlation-TUR to also be proved for transient
dynamics.\vspace{0.5cm}\\
\emph{Acknowledgments.---}
Financial support from Studienstiftung des Deutschen Volkes (to C.\ D.) and the German Research Foundation (DFG) through the Emmy Noether Program GO 2762/1-2 (to A.\ G.) is gratefully acknowledged.

\section*{References}
\bibliography{bib_JPhysA.bib}
\end{document}